# The gap amplification at a "shape resonance" in a superlattice of quantum stripes: a mechanism for high $T_c$

A. Perali[a], A. Bianconi[a], A. Lanzara[a], N. L. Saini[b]

[a] Università di Roma "La Sapienza", Dipartimento di Fisica, 00185 Roma, Italy

[b] Istituto Nazionale di Fisica Nucleare, Dipartimento di Fisica, P. Aldo Moro 2, 00185, Roma, Italy

**Abstract:** The amplification of the superconducting critical temperature $T_c$ from the low temperature range in homogeneous 2D planes ($T_c$<23 K) to the high temperature range (23 K<$T_c$<150 K) in an artificial heterostructure of quantum stripes is calculated. The high $T_c$ is obtained by tuning the chemical potential near the bottom of the *n*th subband ($E_n$), at a "shape resonance", in a range $\mu - E_n \approx \hbar\omega_D$, where $\hbar\omega_D$ is the energy cutoff for the pairing interaction. The resonance for the gap $\Delta_n$ at the *n*th "shape resonance" is studied for a free electron gas in the BCS approximation as a function of the stripe width L, and of the number of electrons ρ per unit surface. An amplification factor $650 > \frac{\Delta_3}{\Delta_\infty} > 6$ for coupling 0.1<λ<0.3 is obtained at the third shape resonance raising the critical temperature in the high $T_c$ range.

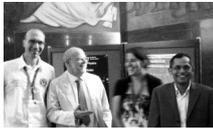
*the authors at Stripes 11 conference Rome July 12, 2011*

## 1. INTRODUCTION

Since the discovery of high $T_c$ superconductivity in a doped cuprate perovskite [1] the search for new materials with higher $T_c$ has advanced by using empirical methods. Most of the proposed theories for high $T_c$ concern new pairing mechanisms but are unable to regulate engineering of new materials with higher $T_c$. Recently a new paradigm for high $T_c$ has been advanced where the critical temperature is proposed to be amplified by a particular heterostructure of the material at the mesoscopic scale length [2-5] making possible the engineering of new artificial high $T_c$ superconducting heterostructures. The heterostructure is made by a superlattice of quantum stripes of size L and period λ. The electronic structure is modified due to **quantum size effects** by reducing L to the mesoscopic range (10-100 Å). **The amplification of the critical temperature is obtained by tuning the Fermi level to the nth "shape resonance"** where $L \approx n\lambda_F/2$, $\lambda_F$ being the Fermi wave length [2-5]. Quantum fluctuations in the quasi-one dimensional (1D) system (L< $\xi_0$) are suppressed by recovering the three dimensional (3D) superconducting phase coherence by Josephson coupling between the superconducting units of the superlattice.

The present paper reports a calculation of the resonant amplification of the superconducting critical temperature at the "shape resonance" in a superlattice of quantum stripes by using the BCS approach. The experimental parameters determined for the cuprate superconductors at optimum doping have been used to calculate the gap as a function of electron number density, stripe width, and coupling factor.

2. THE HETEROSTRUCTURE OF CUPRATE SUPERCONDUCTORS

Let us first describe the standard model of the heterogeneous structure of the Cu perovskite superconductors that is similar to a synthetic heterostructure: a superlattice of superconducting $CuO_2$ layers as shown in Fig. 1a [6-8]. This heterostructure is formed by a plurality of superconducting layers parallel to the (x,y) plane separated by insulating (or metallic) block layers of thickness H. The electrons in the superconducting layers form a two dimensional (2D) electronic system because the effective electron mass $m_z^*$ in the z direction is very large (usually assumed to be infinite) and the separation H between the layers is large enough to neglect single particle hopping along z. The thickness H of the block layers is of the order of the superconducting coherence length in order to allow Josephson coupling between the planes [9,10] giving a 3D superconducting phase below $T_c$ as in synthetic modulated structures [11,12].

The simplest model for the electronic structure of the metallic layers is a two-dimensional electron gas (2DEG) where the one-electron wave function is $\Psi_k(r) = \exp(ik_x \cdot x + ik_y \cdot y)$ and the electron energy dispersion $\varepsilon(k) = (k_x^2/m^* + k_y^2/m^*)\hbar^2/2m_0$, where $m^* = m/m_0$ is the dimensionless effective mass and $m_0$ is the electron mass.

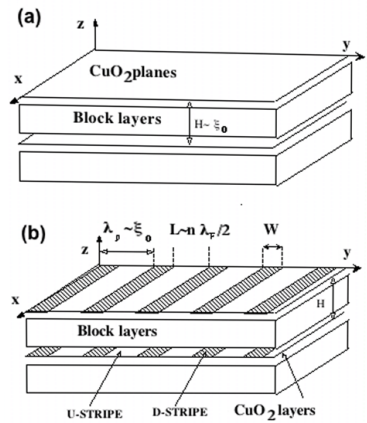

**Fig. 1** Pictorial view of the standard model for the cuprate superconductors formed by a superlattice of Josephson coupled homogeneous $CuO_2$ layers (panel a). Pictorial view of the recently proposed model for the cuprate superconductors formed by a superlattice of quantum stripes in the $CuO_2$ plane (panel b).

The electronic correlation effects are included by taking the effective mass $m^*=2$, a factor 5 larger than in the band structure calculations. The lattice dynamics is characterized by a Debye temperature $\theta_D = \frac{\hbar\omega_D}{K_B} \approx 600\ K$ [10] as experimentally determined for the in-plane Cu-O vibrations. The electron lattice interaction of the homogeneous layer is described by a dimensionless coupling factor $\lambda = N_0 V$, where $N_0 = 2\pi m/h^2$ is the 2D electron density of states and V the strength of the electron-phonon interaction. The superconducting gap is $\Delta_\infty = 2\hbar\omega_D \exp(-1/\lambda)$ and the critical temperature $K_B T_{C\infty} = \Delta_\infty/1.76$ within the standard BCS approach for the coupling factor in the weak coupling range $0.3 < \lambda < 0.1$, where $T_{C\infty} < 23K$.

There is growing experimental evidence that the superconducting $CuO_2$ plane is not homogeneous at a mesoscopic scale. Stripes of undistorted lattice (U-stripes) of width L running in the x direction are intercalated by stripes of distorted lattice (D-stripes) of width W forming a superlattice of stripes with period $\lambda = L+W$, have been found as shown in Fig. 1b. The width L of U-stripes in cuprate perovskites at optimum doping has been measured by different experimental methods [2-5,14-17]. The width L in the $Bi_2Sr_2CaCu_2O_{8+y}$ (Bi-2212) system has been first measured by joint Cu K-edge extended x-ray absorption fine structure (EXAFS) and electron diffraction [2,5,14]. The results have been confirmed by anomalous x-ray diffraction at the Cu K-edge [15]. Recently a more accurate value $L=15\pm0.5$ Å has been determined by joint EXAFS and x-ray diffraction [16]. In $La_{0.185}Sr_{0.15}CuO_4$, a similar superlattice of quantum stripes of undistorted $CuO_6$ octahedra of width L~16 Å has been found [17]. The period $\lambda$ has been determined to be ~25 Å in most of the cuprate superconductors at optimum doping, with $\lambda \sim \xi_0$, satisfying the condition for Josephson coupling in the plane. The width L is quite large and hence the lattice dynamics (mainly the optical phonons and the Debye temperature) and the electronic properties (electron effective mass and the electron-phonon coupling) in the stripes can be considered the same as in the homogeneous 2D layers. In the present model, the separation W between the stripes is large enough to neglect single particle hopping along y but shorter or of the order of the superconducting coherence length in order to allow Josephson coupling in the superconducting phase.

### 3. THE AMPLIFICATION OF THE SUPERCONDUCTING GAP AT A SHAPE RESONANCE

For the particular superlattice of quantum stripes described above the calculation of the superconducting gap at T=0 K is carried out for a single stripe of width L and an infinite potential barrier using the BCS approach. The electron wave vector in the y direction is quantized $k_{ny} = n\pi/L$ and the electronic structure is formed by subbands characterized by the integer n [18]. In the nth subband the electron wave function is $\Psi(r) = \Psi(y)\exp(ik_x \cdot x)$ using the free electron model. The electron dispersion for the nth subband is $\varepsilon_n(k) = (k_{ny}^2/m^* + k_x^2/m^*)(h/2\pi)^2/2m_0 = E_n + \varepsilon(k_x)$ where $E_n = ((h/2\pi)^2/2m)k_{ny}^2$ is the minimum energy for the nth subband where the group velocity of the electrons gets zero. The one-dimensional density of states (1D-DOS) for each subband is $N(\varepsilon) = \dfrac{2 N_0}{\sqrt{2m\varepsilon/\hbar^2}}$, where $N_0$ is the 2D density of states, showing a divergence where the kinetic energy $\varepsilon \to 0$.

The electrons at the Fermi level are called at a "shape resonance" when the Fermi level is tuned close to the bottom of a subband $E_F \sim E_n$ and the Fermi wave vector is $k_F \sim n\pi/L$.

We have solved the BCS equations for the gap parameter and for the chemical potential in a self-consistent manner taking a fixed number of particles. We have followed the approach of Thompson and Blatt [19], who have considered the case of a 2D electron gas confined in a single slab of finite thickness. The electron-electron interaction V, taking into account the interference effects between the wave functions of the pairing electrons in different subbands of the 1D stripe, is written as

$$V(n,j; \varepsilon_n(k), \varepsilon_j(p)) = -V(1+1/2\, \delta_{nj})\, \theta(h/2\pi\, \omega_D - |\varepsilon_n(k)-\mu|)\, \theta(h/2\pi\, \omega_D - |\varepsilon_j(p)-\mu|) \quad (1)$$

where n and j are the subband indices, k and p are the momentum for initial and final states in the pairing process, $\hbar\omega_D$ is the cut off energy, and $\mu$ is the chemical potential. The effective interaction between the electrons in the stripes is characterized by the interference term $(1+1/2\,\delta_{nj})$ due to the superposition of one-electron wavefunctions, that gives higher weight (3V/2) to intraband scattering than to the interband pairing processes with weight V. In the BCS approximation, i. e., a separable kernel, the gap parameter has the same energy cut off $\omega_D$ as the interaction. Therefore in a range $\omega_D$ around the Femi surface it has a value $\Delta_n$ depending only from the subband index and zero outside.

The two self consistent equations for the gap $\Delta_n$ and $\mu$ are:

$$\Delta_n = \frac{V}{L}\sum_{j=1}^{N}\int_{E_1}^{E_2}(1+\frac{1}{2}\delta_{nj})\cdot d\varepsilon'\cdot N(\varepsilon')\frac{\Delta_j}{2\sqrt{(\varepsilon'+\varepsilon_j-\mu)^2+\Delta_j^2}} \quad (2)$$

$$\rho = \frac{1}{L}\sum_{j=1}^{N}\int_{\varepsilon_{\min}}^{\varepsilon_{\max}}d\varepsilon' N(\varepsilon')(1-\frac{\varepsilon'+\varepsilon_j-\mu}{\sqrt{(\varepsilon'+\varepsilon_j-\mu)^2+\Delta_j^2}\theta(\hbar\omega_D-|\varepsilon'+\varepsilon_j-\mu|)}) \quad (3)$$

where L is the stripe width, $E_1 = \max(\varepsilon_{\min}, \mu - \hbar\omega_D - \varepsilon_j)$; $E_2 = \min(\varepsilon_{\max}, \mu + \hbar\omega_D - \varepsilon_j)$; N is the number of the occupied subbands, $\rho$ is the electron density.

These self consistent equations have been solved iteratively, starting with an initial gap parameter equal to a constant and an initial chemical potential equal to the Fermi level in the normal state. The convergence criterion was fixed for differences between the $\mu^{(l)}$ and $\mu^{(l-1)}$, $\Delta^{(l)}$ and $\Delta^{(l-1)}$ less than $10^{-4}$.

We have first calculated the gap for a superlattice of quantum stripes of width L =15 Å [2-5,14-16] and effective mass m*=2 as in cuprate superconductors at optimum doping [20] that is a factor 5 larger than in local density calculation (m*=0.4) neglecting electronic correlation effects [21]. The coupling factor $\lambda$ in the homogeneous plane has been fixed to 0.25 using the calculated value for an homogeneous lattice [21,22] and the Debey temperature has been fixed at 600K. The gap $\Delta_n$ has been calculated as a function of $\rho$ and shown in Fig. 2. The number of electrons per unit surface in the U-stripes of undistorted $CuO_2$ lattice is $\rho = 2(1-\delta_i)/a^2$, where a=5.4 Å and $\delta_i$ is the number of holes per Cu site induced by doping. The maximum density for $\delta_i = 0$ is $\rho_{MAX}= 6.86\cdot 10^{-2}$ Å$^{-2}$, therefore the calculation has been carried out in the range $3<\rho(10^{-2}$ Å$^{-2})<7$ where high $T_c$ superconductivity is expected. The local doping $\delta_i$ in the undistorted stripes is related to the average doping $<\delta_i> = (L/(L+W))\, \delta_i$ [23].

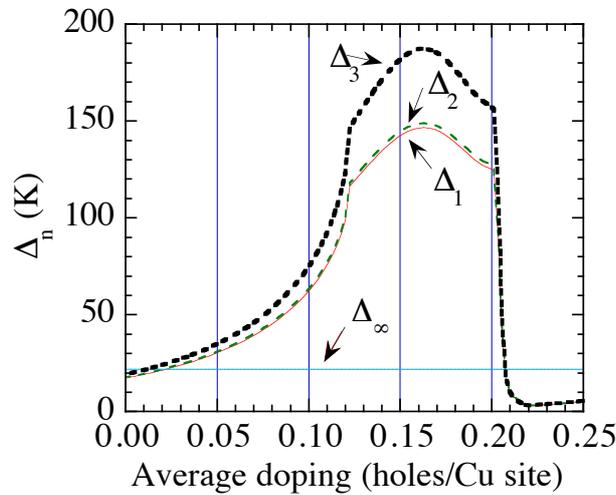

**Fig. 2** The superconducting gaps for a free electron gas with effective mass $m^*=2$, coupling factor $\lambda=0.25$ confined in a superlattice of stripes of width $L=15$ Å as a function of charge density.

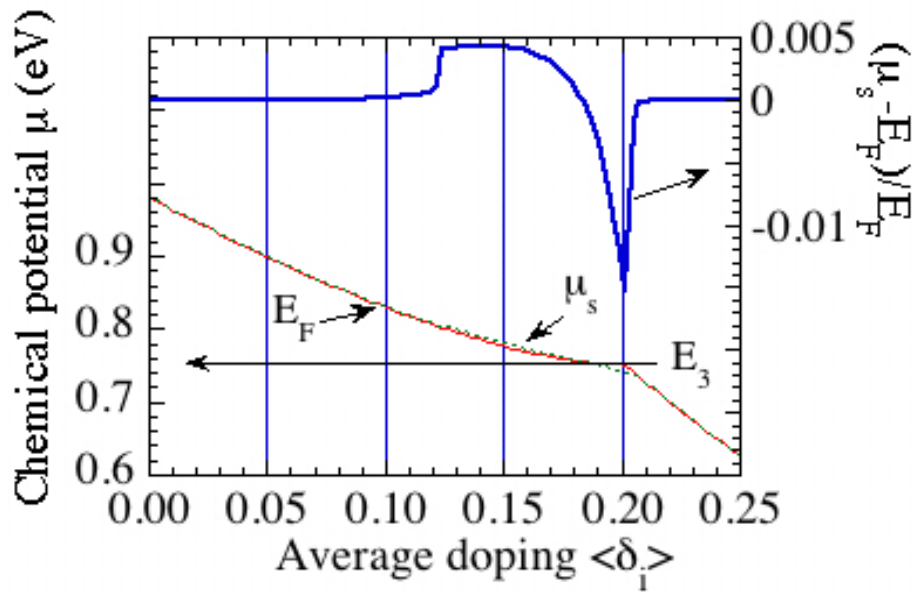

**Fig. 3** The chemical potential in the normal state ($E_F$), in the superconducting phase $\mu_S$ and its relative variation for a free electron gas confined in a superlattice of stripes of width $L=15$ Å as a function of charge density.

In the undoped sample the Fermi level is above the n=3 shape resonance and by doping the chemical potential decreases as shown in Fig. 3 till it reaches the n=3 "shape resonance" where the chemical potential crosses the bottom of the n=3 subband. The superconducting gap shows the resonant amplification when the Fermi level is in the energy range $|E_F - E_3| < \hbar\omega_D$. The maximum of the gap $\Delta_3$ is observed at optimum charge density $\rho = 4.9 \times 10^{-2}$ Å$^{-2}$ or optimum doping $<\delta_i>=0.16$ where the chemical potential $\mu_{opt} \approx E_3 + \hbar\omega_D/2$ that is in qualitative agreement with the experimental value for optimum doping. The width of the resonance covers the range of doping where superconductivity is observed in the phase diagram of cuprate superconductors. The shape of the resonance reproduces the main feature of the phase diagram of $T_C$ versus doping in most of the families of cuprate superconductors with a smooth increase of $T_C$ at low doping, a maximum $T_C$ at optimum doping and decrease in the overdoping regime with a sharp drop. The sharp drop is due to moving of the Fermi energy below the bottom of the n=3 subband.

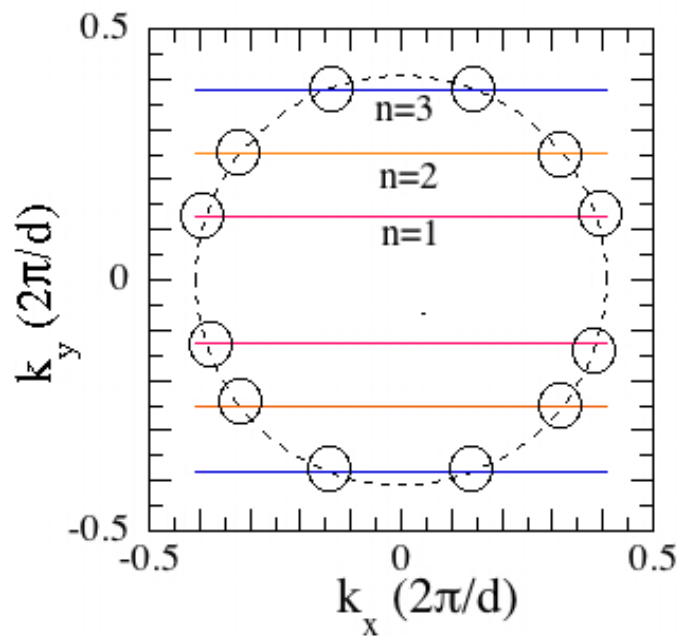

**Fig. 4** The spots (circles) forming the Fermi surface of the three 1D subbands, where the Fermi level is above the bottom of the n=3 subband, for a stripe of width L=15 Å. The electron wave vector is given in units of $2\pi/d$ where d=3.8 Å is the Cu-Cu distance.

Fig. 2 shows that a different gap $\Delta_n$ opens up in each subband. This is due to a non trivial structure in the density of states $N(\varepsilon)$ that introduces an explicit dependence from the subband index in the kernel of the BCS equations. Therefore different gaps $\Delta_n(\mathbf{k}_{Fn})$ will appear for each spot $\mathbf{k}_{Fn}$ at the Fermi surface of the nth subband. At the optimum doping the Fermi surface is characterized by 12 spots as shown in Fig. 4. The gap $\Delta_n$ will appear at the 4 spots of the nth subband. This gives an anisotropic gap structure in the k space and the maximum gap is expected for the upper subband n=3 involved in the "shape resonance". This result is in agreement with experimental findings of anisotropic gap in cuprate superconductors.

The variation of the chemical potential from the normal state ($E_F$) to the superconducting phase $\mu_S$ is shown in Fig. 3. We observe that the superconducting chemical potential $\mu_S$ at the gap maximum is higher than in the normal state ($E_F$) while it is lower where the gap decrease in the overdoped regime. The anomalous large increase of $\Delta\mu/\mu = (\mu_S-E_F)/E_F$ of the order of $3\times10^{-3}$ at optimum doping predicted here is in agreement with experimental data for $YBa_2Cu_3O_{7-\delta}$ [23].

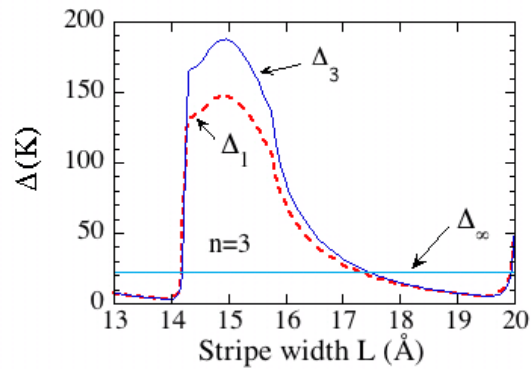

**Fig. 5** The superconducting gaps as a function of the stripe width L in the range of the n=3 shape resonance.

The gap $\Delta_3$ as a function of the stripe width L, with fixed average doping $<\delta_i>=0.16$, is shown in Fig. 5. The amplification of the gap appears in a range of stripe width 14 Å <L< 16 Å that is in good agreement with experimental data. Finally we have calculated the maximum gap $\Delta_3$, with fixed doping and stripe width, as a function of the coupling factor and the amplification factor $A=\Delta_3/\Delta_\infty$ is reported in Fig. 6.

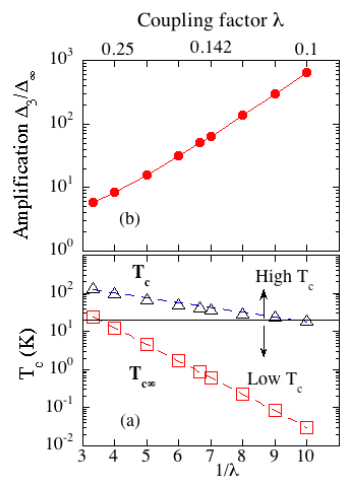

**Fig. 6** The calculated critical temperature $T_{c3}$ at optimum charge density for the n=3 shape resonance compared with the critical temperature $T_{c\infty}$ for the homogeneous plane, in the standard range of coupling factor $0.1<\lambda<0.3$ (panel a). The amplification $A=\Delta_3/\Delta_\infty$ of the superconducting gap $\Delta_3$ in the n=3 shape resonance in comparison with the superconducting gap $\Delta_\infty$ of the homogeneous plane is shown in panel b.

The amplification factor is larger (A~650) for the weak coupling (λ=0.1) and smaller (A~6) for the stronger coupling (λ=0.3). The expected critical temperature at the n=3 shape resonance ($T_c3$) has been estimated by assuming the same amplification factor as for the gap $T_c3 = A\, T_{c\infty}$, where $T_{c\infty}$ is the critical temperature for L->∞. We have plotted critical temperature in the superlattice of quantum stripe with the Fermi level tuned to the n=3 shape resonance, $T_c3$, and for the homogeneous plane $T_{c\infty}$ in Fig. 6 (panel b). $T_{c\infty}$ remains in the low temperature range $0.03 < T_C < 24K$ while by tuning the charge density and/or the stripe width at the shape resonance the critical temperature rises in the high temperature range $20 < T_c3 < 140K$.

### 4. CONCLUSION

By artificial tuning of the charge density and/or the stripe width L at the shape resonance in order to drive the Fermi level near to the $(1/\varepsilon)^{1/2}$ divergence of the 1D DOS, we find a strong amplification of the gap parameter with respect to the bulk value $L \rightarrow \infty$. The amplification $A(\lambda) = \Delta_n/\Delta_\infty$ at the shape resonance is stronger when coupling λ is weaker in the homogeneous superconducting material. The energy width of the resonance in the gap parameter $\Delta_n(\mu)$ depends on the energy cutoff $\hbar\omega_D$. A large value of $\hbar\omega_D$ gives a resonance of the gap $\Delta_n(L,\rho)$ in a wide range of L and/or ρ. The width of the resonance $\Delta_n(\rho)$ has been found in qualitative agreement with the optimum doping range giving high $T_C$ superconductivity in cuprates. The calculated amplification factor for the critical temperature in a superlattice of quantum stripes correctly predicts the rising of the critical temperature from the low $T_c$ to the high $T_c$ temperature range.

The present work points out the main features of $T_c$ amplification by tuning the Fermi level at a shape resonance. We have used first order approximations, such as the free electron model, the BCS theory and infinite barriers. Further work is required to improve the theory. For example in the real systems, presence of a finite potential barrier between the stripes will give a hopping term different from zero resulting broadening of the sharp peaks in the 1D-DOS. This effect is however not expected to be negative if the broadening is of the order of $Ó\omega_D$ while increasing the width of the resonance the $T_c$ amplification is more stable. The introduction of the actual band structure by a tight-binding method is expected to increase the amplification because the separation between the n=3 and n=2 resonances decreases and both are at the shape resonance condition with further enhancement of the critical temperature.

REFERENCES

[1] J. G. Bednorz and K. A. Müller *Rev. Mod. Phys.* **60**, 565 (1988)
[2] A. Bianconi and M. Missori *J. Phys. I (France)* **4**, 361 (1994).
[3] A. Bianconi *Physica C* **235-240**, 269 (1994).
[4] A. Bianconi *Sol. State Commun.* **89**, 933 (1994).
[5] A. Bianconi, M. Missori, N. L. Saini, H. Oyanagi, H. Yamaguchi, D. H. Ha, Y. Nishihara, *J. Superconductivity* **8**, 545 (1995).
[6] J. -M. Triscone, ∅. Fiscer, O. Brunner, L. Antognazza, A. D. Kent, and M.G. Karkut *Phys. Rev. Lett.* **64**, 804 (1990);

[7] Q. Li, X.X. Xi, X.D. Wu, A. Inam, S. Vadlamannati, W.L. McLean, T. Venkatesan, R. Ramesh, D.M. Hwang, J.A. Martinez, and L. Nazar *Phys. Rev. Lett.* **64**, 3086 (1990);

[8] I. Bosovic, J.N. Eckstein, M.E. Klausmeier-Brown, and G. Virshup *Journal of Superconductivity* **5**, 19 (1992)

[9] R. Fastampa, M. Giura, R. Marcon, and E. Silva *Phys. Rev. Lett.* **67**, 1795 (1991)

[10] P. Müller *Festkörperprobleme/Advances in Solid State Physics* **34**, ed. by R. Helbig (Vieweg, Braunschwig) pag.1 (1994).

[11] S. T. Ruggiero and M.P. Beasley in *Synthetic Modulated Structures* edited by L. L. Chang and B.C. Giessen, New York Academic Press (1985) pag. 365;

[12] B. Y. Jin and J. B. Ketterson *Advances in Physics* **38**, 189 (1989)

[13] J. B. Boyce, F. Bridges, T. Claeson, T. H. Geballe, C. W. Chu, and J. M. Tarascon *Phys. Rev. B* **35** 7203 (1987); *Physica Scripta* **37**, 912 (1988); J. M. Tranquada, S. M. Heald, A.R. Moodenbaugh, and M. Suenaga *Phys. Rev. B* **35** 7187 (1987).

[14] A. Bianconi, M. Missori, H. Oyanagi, H. Yamaguchi, D. H. Ha, Y. Nishiara and S. Della Longa *Europhys. Lett.* **31**, 411 (1995).

[15] A. Bianconi, M. Lusignoli, N.L. Saini, P. Bordet, Å. Kvick, P.G. Radaelli *to be published*

[16] A. Bianconi, A. Lanzara, A. Lanzara, A. Perali, T. Rossetti, N.L. Saini, H. Oyanagi, Y. Nishihara, H. Yamaguchi "Spectroscopic Studies of Superconductors" edited by I. Bozovic and D. van der Marel ( SPIE Bellingham 1996)

[17] A. Bianconi, N.L. Saini, A. Lanzara, M. Missori, T. Rossetti, H. Oyanagi, H. Yamaguchi, K.Oka and T. Ito *Phys. Rev. Lett.* to be published

[18] G. Bastard, *Wave Mechanics Applied to Semiconductor Heterostructures* (Les edition de Physique, Les Ulis, France 1988).

[19] J. M. Blatt and C. J. Thompson *Phys. Rev. Lett.* **10**, 332 (1963); C. J. Thompson and J. M. Blatt, *Phys. Lett.* **5**, 6 (1963)

[20] B. O. Wells, Z. X. Shen, D. S. Dessau, W. E. Spicer, D. B. Mitzi, W. L. Lombardo, A. Kapitulnik, and A. J. Arko, *Phys. Rev. B* **46**, 11830 (1992); Z. -X. Shen *J. Phys. Chem. Solids* **53**, 1583 (1992); D. S. Dessau, Z. -X. Shen, D. M. King, D. S. Marshall, L. W. Lombardo, P. H. Dikinson, A. G. Loeser, J. DiCarlo, C. -H. Park, A. Kapitulnik, and W. E. Spicer,*Phys. Rev. Lett.* **71**, 2781 (1993); P. Aebi, J. Osterwalder, P. Schwaller, L. Schlapbach, M. Shimoda, T. Mochiku, and K. Kadowaki *Phys. Rev. Lett.* **72**, 2757 (1994)

[21] O. K. Andersen, O. Jepsen, A. I. Liechtenstein and I. I. Mazin *Phys. Rev. B* **49**, 4145 (1993)

[22] W. E. Pickett, R. E. Cohen, and K. Krakauer *Phys. Rev. Lett.* **67**, 228 (1991); R. E. Cohen, W. E. Pickett, D. Papaconstantopulos and H. Krakauer n *Lattice Effects in High-$T_C$ Superconductors*, edited by Y. Bar-Yam, et al. World Scientific Pub., Singapore, 1992; pag. 223.

[23] A. Bianconi *Sol. State Commun.* **91**, 1 (1994); A. Bianconi, M. Missori *Sol. State Commun.* **91**, 287 (1994).

[24] G. Rietveld, N.Y. Chen, and D. van der Marel *Phys. Rev. Lett.* **69**, 2578 (1992)